\documentstyle[12pt]{article}
\begin{document}
\setlength{\unitlength}{1mm}
{\hfill   January 1996 }

{\hfill    WATPHYS-TH-96-01 }

{\hfill    hep-th/9601154 } \vspace*{2cm} \\
\begin{center}
{\Large\bf
Black hole entropy: statistical mechanics agrees thermodynamics}
\end{center}
\begin{center}
{\large\bf  Sergey N.~Solodukhin$^{\ast}$}
\end{center}
\begin{center}
{\it Department of Physics, University of Waterloo, Waterloo, Ontario N2L 3G1, Canada  \\
and Bogoliubov Laboratory of Theoretical Physics, Joint Institute for
Nuclear Research, Dubna 141980, Russia}
\end{center}
\vspace*{2cm}
\begin{abstract}
We discuss the connection between different entropies introduced for
black hole.
It is demonstrated on the two-dimensional example that the (quantum) thermodynamical
entropy of a hole coincides (including UV-finite terms) with its
 statistical-mechanical entropy calculated according to 't Hooft and regularized
by Pauli-Villars.
\end{abstract}
\begin{center}
{\it PACS number(s): 04.60.+n, 12.25.+e, 97.60.Lf, 11.10.Gh}
\end{center}
\vskip 1cm
\noindent $^{\ast}$ e-mail: sergey@avatar.uwaterloo.ca
\newpage
\baselineskip=.8cm
Since Bekenstein introduced the thermodynamical analogy in black hole physics
\cite{1} and Hawking discovered \cite{2} thermal radiation from a black hole confirming 
this analogy, it is an intriguing problem as to what degrees of freedom are 
counted by the entropy of a black hole.
Equivalently, what (if any) statistical mechanics is responsible for the Bekenstein-Hawking 
entropy? Recently, this problem was attacked from different sides 
and a number of calculations were produced (see reviews in \cite{3}).
Every such calculation typically deals with an specific definition of entropy.
Below I  briefly  list them.

According to 't Hooft \cite{4} the statistical-mechanical entropy $S_{SM}$ arises
 from a thermal bath of quantum fields propagating outside the horizon
 (see also \cite{5}).
An alternative interesting approach  treats entropy as arising from entanglement  
\cite{6}, \cite{7}.
Starting with the pure vacuum state one traces over modes of the quantum field 
propagating inside the horizon and obtains the density matrix $\rho$. 
The entanglement entropy then is defined by the standard
formula $S_{ent}=- Tr \rho \ln \rho$ and is essentially due to correlations 
of modes
propagating at different sides of the horizon. There are formal arguments 
\cite{7}, \cite{8} that $S_{ent}$ coincides with the entropy $S_{con}$ appearing 
in the conical approach \cite{9}-\cite{12}. According to this approach
(see \cite{8}, \cite{12}) one considers the black hole system off-shell by 
fixing only the temperature $T^{-1}=\int g^{1/2}_{00}d\tau$=$ \beta
g^{1/2}_{00}(L)$
on the external boundary and the topology of the black hole geometry. 
An arbitrary metric $g_{\mu\nu}(x)$ satisfying these conditions (the fixing of $T$ 
imposes a boundary condition on the metric) typically 
has a conical singularity with  deficit angle
$\delta=2\pi (1-{\beta \over \beta_H})$ on the horizon. 
The free energy $F[g_{\mu\nu}(x), T ]$ is
then a functional of the metric $g_{\mu\nu}(x)$ inside the region and 
of the temperature $T$ on the boundary. The equilibrium state of the system 
under $T$ fixed is found from the extreme equation $\delta F|_T=0$
and is described by a regular (on-shell) metric with vanishing  deficit angle
($\beta=\beta_H$).
The conical entropy is defined as $S_{con}=- \partial_T F$
where after taking  the derivative $\partial_T$ one considers $S_{con}$ on
 the equilibrium configuration.
The black hole entropy originally appeared within the thermodynamical 
framework and it is determined by total response of the (equilibrium) 
free energy $F$  on variation of temperature: $dF=-S_{TD}dT$.
Remarkably, the conical method gives precisely the thermodynamical entropy, 
$S_{con}=S_{TD}$, for the equilibrium  configuration.

For ordinary thermodynamical systems the thermodynamical and microscopical \\ 
(statistical-mechanical)
entropies are exactly the same. However for the black hole case, 
it was argued in \cite{13}
that black holes provide us with a unique example of a specific system 
for which these entropies
do not necessarily coincide. 

It should be noted that every calculation of statistical entropy encounters  
the problem of dealing with the very peculiar behavior of the 
physical quantities near the
horizon where they typically diverge. To remove these divergences 't Hooft 
introduced ``brick wall'': a fixed  boundary near the horizon within which 
the quantum field does not propagate. Essentially,  this procedure (as it 
formulated in \cite{4}) must be implemented in addition
to the removing of standard ultra-violet divergences. 
Another important point is that this procedure changes the topology of the 
original black hole space-time by introducing an extra 
boundary near the horizon. There are only a few calculations which imply
 ``invariant'' regularization when these extra divergences on the horizon 
are eliminated simultaneously with 
the standard UV divergences. Such  calculations preserve the original 
topology of the black hole space-time. An example is the conical method
 in which, after UV renormalization, 
the residual divergence  of the free energy (at the tip of the cone)
 is proportional to the second power of the  deficit angle and hence does not 
affect the quantities (energy and entropy) calculable for the equilibrium configuration.

The other calculation is a re-formulation of the 't Hooft approach by using 
the Pauli-Villars (PV) regularization scheme \cite{14}. It consists in introducing 
a number of fictitious fields (regulators)
of different statistics and with very large masses. Remarkably, this procedure 
not only yields the standard UV regularization but automatically implements a 
cut-off for the entropy calculation
allowing to remove ``brick wall''.

Both the conical and PV calculations give the ordinary UV divergences for the entropy 
when the regularization is removed. A comparison of the structure of these 
divergent terms\footnote{This concerns only contribution to the entropy
due to the quantum matter minimally coupled to  gravity. In the non-minimal
case the situation is more complicated and at the present time we have no correspondence
between the two methods \cite{8}, \cite{14}.}   for the Reissner-Nordstrom
black hole in four dimensions (see \cite{10} and \cite{14}) shows that they are 
really identical and take precisely the form to provide the correct renormalization 
of bare entropy in agreement with
the suggestion of Susskind and Uglum \cite{9}, \cite{11}. This fact 
suggests that the two calculations really lead to the same result and the 
corresponding entropies $S_{con}~ (S_{TD})$ and $S_{SM}$
are identical including UV-finite terms.

The present state of the theory makes impossible to prove or refute this 
statement in four
dimensions. However,  recent intensive study of  two-dimensional 
models  shows that the black hole physics there looks rather similar 
to the four-dimensional one while the calculations are technically simpler. 
In this note we carry out the precise calculations in two dimensions and 
demonstrate 
equality of the thermodynamical and (regularized by Pauli-Villars) statistical-mechanical
 entropies.

The  black hole thermodynamical entropy at the  one-loop level is 
really a sum 
\begin{equation}
S_{TD}= \left( S_{cl}+ S_{1} \right)[g^{qc}_{\mu\nu}]~~,
\label{1}
\end{equation}
where $S_{cl}$ is entropy coming from the classical gravitational 
action
and $S_{1}$ is a part coming from quantum one-loop term in the 
effective action.
 In principle, both entropies can be calculated off-shell for arbitrary
black hole metric by the conical method assuming that $\beta \neq \beta_H$.  
We get then
the thermodynamical entropy when we set $\beta =\beta_H$ at the end and
the back-reaction is taken into account by calculating the result for 
the quantum-corrected
black hole metric $g^{qc}_{\mu\nu}(x)$ (and $\beta_H^{-1}$ is also assumed to 
be the quantum-corrected
Hawking temperature).  The concrete form of $g^{qc}_{\mu\nu}(x)$ is 
not important for us since all the quantities can be calculated implicitly
 for arbitrary black hole metric.

The form of $S_{cl}$ as function of the black hole geometry
is determined from the classical gravitational action. For concreteness
 we may take it to be some kind of
dilaton gravity:
\begin{equation}
W_{cl}=-\int (e^\Phi R + (\nabla \Phi )^2 +V(\Phi ) ) - \kappa \int R~~,
\label{2}
\end{equation}
where we included also the 2D Einstein term with some ``gravitational
 constant'' $\kappa$. 
Then $S_{cl}=4\pi e^{\Phi (x_+)}+4\pi \kappa$, where $x_+$ is position
 of the horizon.

Essentially, our consideration concerns only that part of $S_{TD}$ which is due to quantum one-loop terms in the action.
We have nothing to say about $S_{cl}$. Its
statistical interpretation requires  additional
 considerations and, possibly, implementation
 of new ideas.

Consider now the quantum massless scalar field described by the action:
\begin{equation}
W_{sc}={1\over 2}\int (\nabla \phi )^2~~.
\label{A}
\end{equation}
In two dimensions it induces the one-loop effective action in the form of the Polyakov term:
\begin{equation}
W_1=-{1\over 96\pi} \int R \Box^{-1} R-{1\over 24\pi}\ln (\Lambda \mu )\int R~~,
\label{AA}
\end{equation}
where we omitted  the 2D cosmological constant which is
 irrelevant to our consideration.
The last term in (\ref{AA}) gives the UV divergence in two dimensions, $\mu$ is
 an appropriate UV-regulator ($\Lambda$
is an infra-red regulator ).
 If we apply Pauli-Villars regularization (see below) then $\mu$ is precisely the
 PV-regulator. This term renormalizes the
``gravitational constant'' $\kappa$. Note that in this model there is no
 renormalization of
$e^\Phi$ which is the real  gravitational coupling in (\ref{2}).

For metric written in the conformal gauge $g_{\mu\nu}=e^{2\sigma}\delta_{\mu\nu}$
 the term
(\ref{AA}) leads to the entropy in the following form \cite{16}
 (see also \cite{10}, \cite{11}):
\begin{equation}
S_{1}={1\over 6}\sigma (x_+)+{1\over 6} \ln (\Lambda \mu )~~.
\label{3}
\end{equation}

Let the black hole instanton be described by the 2D metric:
\begin{equation}
ds^2_{bh}=g(x)d\tau^2+{1\over g(x)}dx^2~~,
\label{B}
\end{equation}
where the metric function $g(x)$ has simple zero in $x=x_+$;  $x_+\leq x \leq L$,
 $0\leq \tau
\leq \beta_H$, $\beta_H={4\pi \over g'(x_+)}$. 
It is easy to see that (\ref{B}) is conformal to the flat disk of radius $z_0$:
\begin{eqnarray}
&&ds^2_{bh}=e^{2\sigma }z_0^2(dz^2+z^2d\tilde{\tau}^2 )~~, \nonumber \\
&&\sigma={1\over 2} \ln g(x)-{2\pi \over \beta_H}\int^x_L {dx\over g}
 +\ln {\beta_H  \over 2\pi z_0}~~, \nonumber \\ 
&&z=e^{{2\pi \over \beta_H}\int^x_L {dx \over g}}~~,
\label{4}
\end{eqnarray}
where
 $\tilde{\tau}={2\pi \tau \over \beta_H}$ ($0\leq \tilde{\tau} \leq2\pi$),
 $0 \leq z\leq 1$. 
So, applying (\ref{3}) we get  for $S_1$:
\begin{equation}
S_1={1\over 12} \int^L_{x_+}{dx \over g}({4\pi \over \beta_H}-g')+
{1\over 6}\ln (\mu \beta_H  g^{1/2}(L)) ~~,
\label{6}
\end{equation}
where we omitted the irrelevant term which is function of $(\Lambda , z_0)$ but not of parameters of the black hole and have retained dependence on UV regulator $\mu$.
The last term in (\ref{6}) really contains logarithm of temperature $T^{-1}=
\beta_H g^{1/2}(L)$ measured on
the external boundary $x=L$. The UV divergence of the entropy (\ref{3}), (\ref{6})
renormalizes the ``gravitational constant'' $\kappa$ in $S_{cl}$ in agreement with
the general statement of \cite{9}, \cite{11}.

Calculate now the statistical-mechanical entropy $S_{SM}$.
Applying Pauli-Villars regularization in two dimensions
one needs to introduce a set of fictitious fields with very large masses:
two anticommuting scalar fields with  mass $\mu_{1,2}=\mu$
and one commuting field with   mass $\mu_3=\sqrt{2}\mu$.
Consider the free energy of the  ensemble of the original scalar field and
 regulators with an inverse temperature $\beta$:
\begin{equation}
\beta F= \sum_n \ln (1-e^{-\beta E_n})
\label{C}
\end{equation}
Note that energy $E_n$ in (\ref{C}) is defined with respect to Killing vector
 $\partial_t~~
(\tau=\imath t )$ and fields are expanded  as $\phi=e^{\imath E t} f(x)$.
 Therefore, $\beta$ in (\ref{C}) is related with temperature $T$ measured at $x=L$
as $T^{-1}=\beta g^{1/2}(L)$. The relevant density matrix is $\rho=\sum_n \phi_n
\phi^*_n e^{-\beta E_n}$, where $\{ \phi_n \}$ is basis of eigen-vectors.
One should take into account that for the regulator fields the Hilbert space hase indefinite
metric and hence a part of regulators contributes with minus sign.

The free energy (\ref{C}) can be determined for the arbitrary
black hole metric (\ref{B}) without reference to the precise form of the metric
 function $g(x)$. 
Repeating the calculation of ref.\cite{14} in this 2D case and applying WKB 
approximation  we finally get 
\begin{equation}
F=-{1\over \pi} \int^\infty_0{dE \over e^{\beta E}-1} \int^L_{x_++h}{dx \over g(x)}
\left( E-2(E^2-\mu^2 g(x))^{1/2} +(E^2-2\mu^2 g(x))^{1/2} \right)~~.
\label{7}
\end{equation}
It should be noted that the WKB approximation for the original massless scalar
 field is really exact.
We introduced in (\ref{7}) a  ``brick wall'' cut-off $h$. In fact, one can
 see that divergences at small $h$
are precisely cancelled in (\ref{7}) between the original scalar and the
 regulator fields.
This is 2D analog of the mechanism discovered in \cite{14}.
So one can remove the cut-off in (\ref{7}). However we will keep it  
arbitrarily small in the process of calculation of
 separate terms entering in (\ref{7}).

It is straightforward to compute the contribution of the original massless field in
 (\ref{7}). For computation of the regulator's contribution take  the 
fixed $E$ and consider the integral:
\begin{equation}
I[\mu ]= \int^{L_E}_{x_++h}{dx \over g(x)}(E^2-\mu^2 g(x))^{1/2}~~,
\label{8} 
\end{equation}
where integration is doing from the horizon ($x_++h$) to distance $L_E$ defined from equation
$g(L_E)={E^2 \over \mu^2}$.  It is clear that when $\mu$  grows
$L_E$ becomes closer and closer to ($x_++h$). So, considering 
 limit of large $\mu$ we conclude that integral (\ref{8}) is concentrated
near the horizon where we have: $g(x)={4\pi \over \beta_H}(x-x_+)=({2\pi \rho \over
\beta_H})^2,~~{dx\over g}={\beta_H \over 2\pi}{d\rho \over \rho}$ and the new
radial variable $\rho$ now runs from  $\epsilon=\sqrt{\beta_H h \over \pi}$ to $ ({E\beta_H \over 2\pi \mu})$.
The integral (\ref{8}) then  reads:
\begin{eqnarray}
&&I[\mu ]=\mu \int_\epsilon^{E\beta_H \over 2\pi \mu}{d \rho \over \rho} \sqrt{({E\beta_H \over 2\pi \mu})^2-\rho^2} \nonumber \\
&&={(E\beta_H) \over 2\pi} \left( arctanh \sqrt{1-({\mu \epsilon 2\pi \over
 E \beta_H})^2}-\sqrt{1-({2\pi\epsilon\mu \over E \beta_H})^2} \right)~~.
\label{9}
\end{eqnarray}
Using the  asymptote $arctanh (1-x)=-{1\over 2} \ln {x\over 2}+ o(x)$ we finally get
\begin{equation}
I[\mu ]=-{(E\beta_H) \over 2\pi}(1+{1\over 2} \ln 2 +\ln ({\mu\epsilon\pi \over E \beta_H}))~~.
\label{10}
\end{equation}
This is the key identity allowing computation of the free energy (\ref{7}).
Omitting details which are rather simple below is the result:
\begin{eqnarray}
&&F=-{1\over 12}[ {\beta_H \over 2 \beta^2} \int^L_{x_+}{dx \over g}
({4\pi \over \beta_H} -g')+{\beta_H \over \beta^2} \ln ( \mu \beta g^{1/2}(L))]
 \nonumber \\
&&+{\beta_H \over \beta^2}[{1\over 2\pi^2} \int^\infty_0{dxx\ln x \over e^x-1}-
{1\over 12}(1-\ln 2) ]~~,
\label{11}
\end{eqnarray}
where we removed the brick wall cut-off and used that $
\int^\infty_0{dx x \over e^x-1}={\pi^2 \over 6}~~.$  The statistical-mechanical free energy (\ref{11}) is really an off-shell quantity (see \cite{14}) defined for  
arbitrary metric (\ref{B}) and $\beta$ not necessarily equal to $\beta_H$.

Calculating now entropy $S_{ST}=\beta^2\partial_\beta F$ and putting $\beta=\beta_H$
we obtain:
\begin{equation}
S_{ST}={1\over 12} \int^L_{x_+}{dx \over g}({4\pi\over \beta_H}-g')+ {1\over 6}\ln (\mu \beta_H g^{1/2}(L)) +C~~,
\label{12}
\end{equation}
where $C$ is some numerical constant not depending on $\mu$ or metric $g(x)$.
So, we see that $S_{ST}$ exactly coincides with $S_1$ (\ref{6}).
We conclude that at least this part of the thermodynamical entropy has the 
statistical meaning. 

Various calculations of black hole entropy in two dimensions were recently considered
 in \cite{17}. In particular,
it was concluded that $S_{SM}$ and $S_{TD}$ are different: one has to subtract
 a (divergent)
contribution of the fictitious Rindler particles from $S_{SM}$ in order to get $S_{TD}$.

The correspondence  of our results with that of \cite{17} is seen from analysis
of important interplay of two different limits $h\rightarrow 0$ (brick wall)
and $\mu^{-1}\rightarrow 0$ (UV-regulatorization). If one takes the limit
$\mu^{-1}\rightarrow$ first one obtains that contribution of the regulators in the
free energy (10) completely vanishes. On then gets the quantities which are functions of the brick wall parameter $h$ and divergent in the limit $h\rightarrow 0$.
These are that quantities calculated in \cite{17}. Elimination of their divergence (with respect to limit $h\rightarrow 0$) might require some subtraction procedure proposed in 
\cite{17}. Note, that in this regime the "brick wall" is treated as real boundary staying at
{\it macroscopical} distance $h$ from the horizon with $h$ being larger than any
UV cut-off $\mu^{-1}$.

The situation is different if we consider "brick wall" as an fictitious imaginary boundary
with $h$ being smaller any scale $\mu^{-1}$ of UV cut-off. Then the "brick wall"
divergences are eliminated by the standard UV-regularization and the
UV-regulators do contribute to the free energy and entropy\footnote{This probably
means that we do not need in "brick wall" at all ($h=0$ from very beginning) in order to
formulate the statistical mechanics of quantum fields around black hole.
We must just impose some analyticity condition on quantum field wave functions
at the horizon and deal with continuous energy spectrum of the quantum
field system.}. This contribution is concentrated at the horizon. it leads to appearance of
additional terms in the entropy (15) that are finite after renormalization. It is worth noting that mechanism of this phenomenon  is similar to that of the conformal anomaly.
This similarity is not occasional since the result for the statistical entropy
(15) occurs to coincide with the thermodynamical expression (5), (8) which is indeed 
originated from the conformal anomaly of the effective action (4).

We do not have this phenomenon in the statistical mechanics on space-time without
horizons where the statistical entropy was proved to be conformal invariant and not
dependent on UV cut-off (see \cite{D}). This is easily seen from our analysis.
Indeed, in this case we have $g(x) \geq g_0>0$ everywhere and for large
UV cut-off $\mu> \mu_0={E\over g^{1/2}_0}$ contribution of the regulators
disappears in the free energy (10).

Thus, in the presence of horizons the statistical mechanics of quantum fields depends
on their UV behavior. The UV-regulators lead to non-trivial contribution to
statistical entropy that is finite after renormalization. Unfortunately,
the straightforward generalization of this result on higher dimensions
meets the still open problem of statistical description of the non-minimally
coupled conformal matter \cite{16}.

Concluding, we propose that there is some unification: all the entropies ($S_{SM}$,
  $S_{ent}$, $S_{con}$, $S_{TD}$) introduced for a black hole which arise
 due to quantum minimal matter
are really identical. 
In two dimensions this statement can be supported by precise calculation. Recall,
 that this does not concern the classical entropy of a hole for the statistical 
explanation of which a special consideration is required.

{\bf Acknowledgments:}  I am grateful to Robert Myers for encouraging
me to prepare this calculation for publication. I also thank Robert Mann
for valuable comments. 
This research was supported by NATO Fellowship and in part by the Natural
Sciences and Engineering Research Council of Canada.

\end{document}